\documentclass[12pt]{article}
\usepackage{amssymb}

\begin{document}

\pagestyle{empty}

\hspace*{2.0in}
\begin{center}
\renewcommand{\baselinestretch}{1.2}
\large   
{\textbf{ON THE
FOUR-COLOR-MAP THEOREM}}
\end{center}

\vfill

\renewcommand{\baselinestretch}{1.0}
\normalsize   
\begin{center}

\renewcommand{\baselinestretch}{1.0}
\normalsize   
\begin{center}
A. Petermann\\ CERN\\ CH - 1211
Geneva 23, Switzerland
\end{center}

\author{A. Petermann\\ CERN\\ CH - 1211 
Geneva 23, Switzerland}
\end{center}

\vfill

\begin{abstract} Coloring 
planar Feynman diagrams in spinor quantum
electrodynamics, is a non 
trivial model soluble without computer. Four colors are
necessary and 
sufficient. 
\end{abstract}

\vfill

\begin{flushleft} CERN\\Revised version July 2004 \end{flushleft}

\vspace{2.7in}

\pagestyle{plain}
\newpage

\section*{INTRODUCTION}

The representation of 
fundamental processes in the theory of elementary particles
proceeds 
mainly by means of the so-called Feynman diagrams. this is the case, for
instance, for the gauge theories (QED, Weak interactions, QCD), Higgs 
scalars, and
others. These diagrams, as is well-known, are drawn on a plane, on which different
kinds of lines represent the various elementary particles entering the
interactions. Therefore (see Appendix), they are similar to the maps of 
different
countries which have given rise to the well-known four-color map theorem
(K.~Appel
\& W.~Haken, 1977)~\cite{Appel}.

We shall say that ``normal maps'', as defined by
Kempe~\cite{Kempe}, summarize the set of rules
in terms of vertices, planarity, boundaries of
adjacent countries, number of neighbors and
so on. Coloring those countries, demands that
two countries with an adjacent border must
have different colors. And the fact that four
colors are always enough to color any normal
map is the Appel-Haken four-color  theorem.

In the present paper we shall consider the case of countries that represent a
planar self-energy of an electron, in QED with any number of photons emitted and
then reabsorbed, forming the photon cloud of the physical electron. This is a
spinor quantum electro-dynamics problem. 
The rules are, for the most important: energy conservation, electric 
charge conservation and Furry's theorem (which stems from the properties 
of Clifford algebra. In other words from the algebra of a Dirac 
gamma-matrices in four dimensions). The grammar is more restrictive, and 
therefore simplifies the problem of coloring the countries, the borders of 
which are both photon and electron lines (resp. wavy and full lines).

In a first section, we shall examine the case when the electron line only emits and reabsorbs photons in a planar way. This fundamental case will be called the
``rainbow case''. The various colors will refer to the numbers: 1, 2,
3, 4. It will be shown that such a configuration needs only four
colors for the represented countries. Then in a second section, we
consider the inclusions of fermion loops: vacuum polarization,
scattering of light by light, generally electron--positron boxes
with interaction of any even number of photons. It will be shown
that those inclusions will not destroy the result of the first
section, the rainbow case. We conclude therefore that the four-color
map theorem for coloring the regions of an electron self-energy planar
diagram in spinor quantum electrodynamics is confirmed without the use
of a computer.

\subsection*{Addendum to the Introduction:} As a warm up exercise, the interested
reader can try to color diagrams of the planar $\varphi^4$-theory, and find that
only two colors are enough in order to color the regions of any diagrams.

\section{RAINBOW CASE}

We start by dealing with the electron self-energy case in QED planar theory. That
is, the electron emits $n$ photons which are all reabsorbed when reaching the
final state. Planarity means that photons-lines do not cross and we
are therefore in the so-called `rainbow' case. We shall not enter in
this note into the technicalities of rainbow diagrams. Several
exhaustive investigations have been made in various cases~\cite{Hooft}.

We draw an horizontal fermion line (full line) and cross this line vertically by a
cut in its middle. If a total of $n$ `photons' lines (wavy lines), are exchanged
from one side of the cut to the other side and if the cut is `maximal' (in other
words if all emitted photons cross the cut before being reabsorbed), there will be
$C^{n}_{n-\kappa}$ combinations for the emission of $n-\kappa$ lines in the upperhalf plane and $\kappa$ lines in the lower half plane. Similarly,
$C^{n}_{n-\kappa}$ ways of absorbing these photons lines.

We call this the `maximal rainbows' at order $n$. As it is well known in the theory
of Feynman diagrams, there are also rainbow of order $n$ ($n$ wavy lines) which are
not maximal. They have, along the electron line, between the emission or
absorption of two photons lines of the main rainbow, on each side of the cut,
some so called `subrainbows' with $\ell$ photons lines. These subrainbows have the
same properties as the maximals but are of the order $\ell$ . To analyse them, if
it is not evident to the reader, one can also define subcuts for them and the
results follows with $\ell$ instead of $n$ lines. The combinatorics, in those
case, is trivial and we leave the alert reader to verify it.\footnote{The case of
a subrainbow connected to another subrainbow only by the electron line and no
photon line is not considered. By cutting the electron line between the rainbows
one is led to two disconnected diagrams.}

The above consideration allow ourselves to formulate the following theorem:

\noindent{\underline Theorem: } In the pure rainbow case (no electron loops), for
every planar configuration, at any $n$, the partition of colors can be the
following: colors 1 and 3 for the upper half-plane; 2 and 4 for the lower half
plane. \vglue.5cm
If follows that four colors are always enough to color the rainbow configurations
of an electron self-energy. Two adjacent regions will always have borders with
different colors on each side. The trick is to start with color 1 upside the most upper photon line. Then alternating with color 3, once the photon boundary of
this upper region is crossed. Since photon lines do not cross (planarity), one
always ends with colors 1 or 3 along the electron line boundary.

\section{THE INSERTIONS OF PLANAR FERMION BOXES IN THE RAINBOW DIAGRAMS}

The rainbows diagrams dealt with in Section 1 are subject of corrections due to
the insertions of planar electron boxes (ex. vacuum polarization, scattering of
light by light, hexagons ...). The electron lines of these boxes are in turn,
corrected by rainbows themselves containing boxes and so on until the maximum
order is reached, order which can tend to infinity. Finding a rule of coloration
for all kind of boxes inserted in a rainbow diagram, say the maximal rainbow at
order  $s < n$, such that this corrected rainbow is of order $n$, will give the
rule for the corrections (photons propagators, boxes of all sizes) to the boxes
insertions themselves by recursion.

This rule of coloration stems fundamentally from a property of Clifford algebra in
four-dimensions: the algebra of Dirac $\gamma$-matrices. For each electron box,
corresponds a Trace of the product of $\gamma$-matrices. It turns out to be
identically zero, if the number of $\gamma$'s is odd. Therefore, only boxes with
an even number of sides survive. This is all we need in order to show that boxes'
insertions in any simple rainbow diagrams do not need more then four colors.
Indeed, suppose we are dealing with a rainbow in the lower half-plane. The boxeshaving only an even number of neighbors, if we alternate for this rainbow the
colors 2 and 4, a single box insertion, with color 1 and inside photon corrections
with colors 1 and 3 alternating, will provide, trivially, configurations with no
two adjacent neighbors with the same color. The box internal photon rainbows
corrections, are of colors 1 and 3, whereas the external photon rainbows
corrections to the box will be of colors 2 and 4 (the same alternating colors as
the lower half-plane rainbows). Finally the insertion of further (or several)
box(es) inside the box-corrections themselves, will alternate with the other pair
of colors 2 and 4 and so on and so forth\footnote{It is easy to notice the
self-similarity of the corrections at all scales, from which is derived the
recursive procedure. By all scales, we mean dominant, sub-dominant,
sub-sub-dominant and so on (see also Section 3).}. All the previous properties aredue to the main fact that no boxes with odd neighbors will ever appear. This fact
spots the crucial difference between the general geographic case of Appel and
Haken, and the planar Feynman diagrams for quantum electrodynamics. It explains
that the constraints of an additional grammar-rule (no odd neighbors countries to boxes) to the general `normal case' for drawing countries, makes the problem so
considerably simpler, that it can be dealt with, without recourse to the computer's
aid. The way used to color any planar spinor Q.E.D. self-energy diagrams uses a
\underline{systematic} alternance of pairs of colors (1, 3) and (2, 4) and
provides the rule of coloring we were looking for. That is:

\noindent{\underline{Theorem:}} four colors are sufficient in order to color any
planar electron self-energy diagram in spinor quantum electrodynamics.

\section{CONCLUSIONS AND REMARKS}

Why only planar graphs have been considered in this paper is self-evident.
Non-planar diagrams have indeed parts in common in the plane and are therefore
meaning less for the color-problem we have investigated. The restriction to planar
diagrams (here the self-energy) does nevertheless have a link with physical
situations. An example is discussed
in~\cite{Brezin} for a $(\vec\phi^2)^2$ field
theory with O(N) symmetry. One sees that, in
the calculations of the critical exponents,
the leading and subheading terms in a 1/N
expansion are all planar. The same happens
when considering a field theory\cite{Wilson},
involving one scalar field an N
massless Dirac fermion-fields $\psi^i$ and
$\vec \psi^i$ coupled via a Yukawa type interaction. And then performing a 1/N
expansion.

Here, however, our aim was simply to figure out a non trivial example where maps
(here Feynman diagrams) need four colors and are sufficient to avoid two
countries with the same color on each side of the border which separates them. As
emphasized, the grammar for the drawing of countries is more restrictive:

\begin{itemize}
\item[--] No odd-sized fermion polygons; 
\item[--] Two kinds of borders: fermions (full lines) and
photons (wavy lines); 
\item[--] All vertices of order three, with two fermion-lines and one
photon lines ending at these vertices.
\end{itemize}

The proof proceeds by recursion, since the same patterns of diagrams, reduced in
scale, always appear at a lower level, showing a patent self-similarity.
Physically this can be translated in the following way: each particle has a
virtual accompanying cloud which surrounds it as  it propagates. And each of the
virtual particles in the cloud also drags along, its own virtual cloud and so on
ad infinitum. Paraphrasing R.
P{\'e}ter~\cite{Peter}, recursion is based on
the same thing happening on several different
levels at once.

\newpage
\vfill
\begin{center}
\vglue.3cm 
\large{\textbf{Appendix}}
\end{center}
As we said in the Introduction, in physics (especially in field theory) it 
is well known that the representation of interactions between 
elementary particles can be visualized by means of the so-called Feynman 
diagrams. For a given process, described by Green functions, the usual way 
of calculating its contribution is to use perturbation expansions in terms 
of a given interaction density Lagrangian, which characterizes the theory 
one is dealing with. One of the most popular examples is spinor Quantum 
Electrodynamics (QED for short) for which the Lagrangian is ~eu*Au with u 
the electron field and A the photon field vector product with Dirac 
matrices (4-dimensional). 
\linebreak
If one wants to draw all conceivable Feynman diagrams corresponding to a 
given process to the perturbative order n, the standard method is the 
derivation of all these diagrams from a generating functional W(Ji) (Ji 
being the sources of the various particles)(Ref.~\cite{Schwinger}). W(Ji) 
is obtained as the logarithm of the partition functional Z(Ji). This 
method is described with full details in all quantum field theory 
textbooks, so we shall not develop it here. Our main interest in this note 
is coloring planar Feynman diagrams according to Tait's method of coloring 
edges of a diagram. This means that at each cubic vertex corresponding to 
a given process to the perturbative order n, the three edges segments 
starting from this vertex are of different colors, three colors having 
been fixed once for all (Ref.\cite{Tait}) . The generating functional, 
when expanded in powers of the coupling (e in the case of QED), will give 
rise, as already said, to all possible cubic diagrams at each order n. But 
here, we are, for evident reasons, only interested in planar diagrams of 
which we want to color the facets (countries). So we must first proceed to 
a filtering and eliminate a plethora of non-planar and improper diagrams 
which are of no interest for our scope. The residue of this filtering will 
provide all conceivable diagrams and can be shown to be topologically 
equivalent to all conceivable geographical cubic maps. By Wick ordering 
and contractions (contractions give rise to propagators (=edges)), and 
since at each vertex three different colors meet, all the edges are 
3-colored according to Tait's coloring; and from this, the Petersen 
conjecture on hamiltonian paths and disjoint even subcircuits 
(Ref.\cite{Petersen}) is also demonstrated. This conjecture is equivalent 
to the 4-color problem. This proof has already been done in the case of 
QED, using straightforward methods. But according to the present 
note, is valid for any cubic geographical map. In conclusion, the Feynman 
diagrams of QED have the crucial advantage, over other methods of coloring 
maps, that one can always follow the path of the electric charge, which is 
a conserved quantity. Thus, deleting all wavy lines (photons), one is 
dealing with the fermionic motion of the charge, which is a hamiltonian 
circuit or a collection of mutually disjoint even subcircuits. And this is the Petersen condition for the 4-color map theorem to hold. 
 
\end{document}